\documentstyle[12pt]{article}
\begin{document}

\def\advphys{{\it Adv. Phys.}}
  \def\annphys{{\it Ann. Phys.~}}
  \def\jltp{{\it J.~Low Temp. Phys.~}}
  \def\prl{{\it Phys. Rev. Lett.~}}
  \def\pr{{\it Phys. Rev.~}}
  \def\pra{{\it Phys. Rev.~A~}}
  \def\prb{{\it Phys. Rev.~B~}}
  \def\prc{{\it Phys. Rev.~C~}}
  \def\prd{{\it Phys. Rev.~D~}}
  \def\rmp{{\it Rev. Mod. Phys.~}}
  \def\zetf{{\it ZhETF~}}
  \def\jetp{{\it  JETP~}}
  \def\jetpnew{{\it  JETP~}}
  \def\pisma{{\it Pis'ma ZhETF~}}
  \def\jetpl{{\it  JETP Lett.~}}
  \def\pletta{{\it Phys. Lett.~A}}
  \def\plettb{{\it Phys. Lett.~B}}
  \def\epl{{\it Europhys. Lett.~}}
  \def\jpa{{\it J. Phys. A~}}
  \def\jpb{{\it J. Phys. B~}}
  \def\jpc{{\it J. Phys. C~}}
  \def\jpcm{{\it J. Phys.: Condensed Matter~}}
  \def\jpd{{\it J. Phys. D~}}
  \def\jpe{{\it J. Phys. E~}}
  \def\jcond{{\it J. Phys.: Cond. Matter~}}
  \def\pscr{{\it Phys. Scr.~}}
  \def\npa{{\it Nucl. Phys.~A~}}
  \def\npb{{\it Nucl. Phys.~B~}}
  \def\ptp{{\it Prog. Theor. Phys.~}}
  \def\prep{{\it Phys. Rep.~}}
  \def\jphysique{{\it J. de Physique~}}
  \def\jphysqlett{{\it J. Phys. (Paris) Lett.~}}
  \def\zphys{{\it Z. Physik~}}
  \def\zphysa{{\it Z. Physik A~}}
  \def\zphysb{{\it Z. Physik B~}}
  \def\zphysc{{\it Z. Physik C~}}
  \def\zphysd{{\it Z. Physik D~}}
  \def\physica{{\it Physica~}}
  \def\physicaa{{\it Physica A~}}
  \def\physicab{{\it Physica B~}}
  \def\physicabc{{\it Physica B+C~}}
  \def\physicac{{\it Physica C~}}
  \def\physicad{{\it Physica D~}}

\begin{center}
{\large{\bf {
Fermions on quantized vortices in superfluids and superconductors
}}\\

\vskip 0.5 truecmLecture Notes \\
 Summer School on Condensed Matter Physics: \\
{\it Symmetry of
the Order  Parameter in High Temperature Superconductors}\\
  Ankara
17-22 June 1996.
}

\vskip 0.5 truecm

\obeylines{G.E. Volovik\footnote{volovik@boojum.hut.fi}
\vskip 0.5 truecm
Low Temperature Laboratory,
Helsinki University of Technology
Otakaari 3A, 02150 Espoo, Finland
and
L.D. Landau Institute for Theoretical Physics,
Kosygin Str. 2, 117940 Moscow, Russia
}
\end{center}

\vskip 1 truecm

\begin{abstract}
The bound states of fermions in cores of quantized vortices in
superconductors and Fermi superfluids and their influence on the
vortex dynamics  are discussed. The role of the spectral flow of the
fermions through the gap nodes is emphasized.
\end{abstract}

\vfill \eject

\tableofcontents

\vfill
\eject

\section{Introduction}

 Gapless fermions govern the low temperature behavior of
condensed matter. In homogeneous condensed matter the distinctive feature of
the spectrum of the gapless fermions is the dimension of  zeroes in the
spectrum: the dimension $D$ of the manifold in 3-dimensional momentum space,
where the energy of the fermion vanishes: $E({\bf p})=0$. There are $D=0$ point
node, $D=1$ line  of zeroes, $D=2$ Fermi surface  and $D=3$ Fermi
band  (the fermionic condensate -- a flat plateau in the
quasiparticle energy at the Fermi level \cite{Khodel1990,Khodel1994}).
The topological objects
lead to the zeroes of the gap in the real space (say, on the vortex axis).

At the moment several groups of  experiments support  the possibility of
lines of
of nodes in the electron spectrum of some of the high-T$_c$ materials (see
Reviews \cite{Scalapino1995,Annett1996,vanHarlingen1995}. The point nodes
result in many interesting properties of the superfluid $^3$He-A
\cite{Volovik1992a}. The gapless fermions concentrated near the vortex axis
are important for the low-temperature dynamics of vortices in all
superconductors and Fermi superfluids. Here we discuss some consequences of
the gap nodes in bulk superconductors and in the core of vortices.

\section{Electron density of states due to the gap nodes.}

\subsection{DOS in bulk superconductors and superfluids.}

The gap nodes in the superconducting gap lead to the power law dependence of
the density of states (DOS) on the frequency $\omega$:
$$ N(\omega)=2\int {{d^3p}\over {(2\pi)^3}}~ \delta (E({\bf
p})-\omega) \sim N_F \left( \omega\over \Delta \right)^\alpha~~,
\eqno(2.1)$$
where $N_F$ is the DOS of the normal state,  $E
=\sqrt{\varepsilon^2+\vert\Delta({\bf p})\vert^2}$ is the quasiparticle
excitation spectrum. The exponent $\alpha=2-D$, ie  $\alpha=2$ for the
point nodes,
$\alpha=1$ for lines of nodes and $\alpha=0$ if the Fermi surface remains in
the superconducting state.

Situation changes in the presence of
impurities: eg intermediate values $1>\alpha>0$ are possible for the case of
the lines of gap nodes in the presence of impurities \cite{Nersesyan1994}.

\subsection{DOS induced by supercurrent and texture.}

The gap nodes in the presences of textures lead to the nonzero  density
of states at zero energy $N(0)\neq 0$. Texture is a slow space variation of the
order parameter: in superconductors due to the crystal lattice the variation
of the phase of the order parameter is only allowed, which gives the
supercurrent with the superfluid velocity
${\bf v}_s= (\hbar/2m)(\vec \nabla \Phi -e{\bf A})$.
In superfluid $^3$He the anisotropy axis ${\hat {\bf l}}$ can vary slowly in
space (in $^3$He-A the unit vector ${\hat {\bf l}}$ shows also the direction to
the point gap node in momentum space). The DOS in supercurrent can be obtained
from Eq.(2.1) by substitution
$\omega \rightarrow p_F v_s$. In ${\hat {\bf l}}$-texture one should
substitute
$\omega^2 \rightarrow v_F\Delta  \vert \nabla {\hat {\bf l}} \vert  $.
As a result one has
$$ N(0) \sim N_F \left(  v_F \vert \nabla {\hat {\bf l}} \vert  \over \Delta
\right)^{\alpha/2}~~,
\eqno(2.2)$$
for the texture (see \cite{VolovikMineev1981} for $\alpha=2$)
and
$$ N(0) \sim N_F \left( p_Fv_s\over \Delta \right)^\alpha~~,
\eqno(2.3)$$
for supercurrent (see  \cite{Nagai1984} for point nodes, where $\alpha=2$, and
\cite{Muzikar1983} for lines of zeroes, where  $\alpha=1$).    Thus the gap
nodes in the presence of texture or supercurrent give  the linear term in
heat capacity $C\propto  N(0) T$ at low $T$.

The physical origin of the nonzero density of states due to supercurrent is the
same as that of the depairing current in a conventional superconductor.  In the
presence of a supercurrent the quasiparticle excitation spectrum is
shifted by an amount ${\bf p}\cdot{\bf v}_s$.
The density of states at the Fermi level is given by
$$N(0)
=2\int {{d^3p}\over {(2\pi)^3}}~ \delta (E({\bf
p})+  {\bf p}\cdot{\bf
v}_s)                        ~~.\eqno(2.4)$$
For the superconductor with nodes, the shift transforms the
nodes into the Fermi surfaces, at which the quasiparticle energy
$E+ {\bf p}\cdot{\bf v}_s$ is zero. The Fermi
surface   gives the finite local DOS
$N(0)\propto \vert{\bf p}_0\cdot {\bf v}_s\vert^{\alpha}$, where
${\bf p}_0$ is the direction to the node.

The dependence of the density of states on the supercurrent leads to
the nonlinear Meissner effect \cite{Xu1995} and to the power-law
dependence of heat capacity on the magetic field in the mixed state
of superconductors (see next Section).

\subsection{Magnetic-field dependence of DOS.}

The  DOS averaged over the elementary cell area $A_{\rm cell}$
of the  Abrikosov vortex lattice is
$$N(0) = {1\over A_{\rm cell}}\int_{A_{\rm cell}}
d^2r N(0,{\bf r}) ~\sim N_F {1\over A_{\rm cell}}\int_{A_{\rm cell}}
d^2r \left( p_Fv_s({\bf r})\over \Delta
\right)^\alpha~~.\eqno(2.5)$$
Since $v_s \propto 1/r$ outside the core of the vortices, the integral
diverges. The main contribution to the integral comes from the region far
from the
vortex core, at the distance of order the intervortex spacing
$\sqrt{A_{\rm cell}}\propto 1/\sqrt{H}$, which serves as the upper cutoff in
the integral. This gives the following DOS in the mixed state of the
superconductors with gap nodes:
$$N(0)\propto N_F\left( {H\over H_{c2}}\right)^{\alpha/2} ~~.\eqno(2.6)$$
For a superconducting state with lines of nodes ($\alpha=2$) the dominant
magnetic-field dependence is $N(0)\sim N_F \sqrt{H/H_{c2}}$, while in
superconductors with point nodes $N(0)\sim
N_F  H/H_{c2} $ or $N(0)\sim N_F  H/H_{c2} \ln (H_{c2}/H) $  depending on the
orientation of the point nodes with respect  to the
magnetic field \cite{Volovik1988}. Recent measurements of the
magnetic-field  dependence of the heat capacity
\cite{Moler1994,Moler1995,Fisher1995} are consistent with the square-root
law (see also \cite{Ramirez1995} for the heavy fermionic compound
UPt$_3$). Note that the impurities decrease the exponent $\alpha$
\cite{Nersesyan1994}.

\subsection{Topological stability of nodes.}

 Point nodes and Fermi surfaces are topologically stable: they are described by
integer topological invariants (charges) expressed in terms of Green's
function \cite{Volovik1991,Volovik1992a}. Those point nodes and Fermi
surfaces, which have nonzero topological charge, are robust: they  survive
under external perturbations and can disappear only by mutual annihilation
with zeroes of the opposite charge. Of course, the conservation of the
topological charges does not exclude the possibility of transformation of
zeroes to that of higher  dimension: the point node can transform into the
closed fermi surface, in turn the fermi surface can transform into the fermi
band \cite{Khodel1990,Khodel1994,Nozieres1992}. The latter represents the
fermi surface of finite thickness or  a pair of half-quantum vortices in
momentum space, which is decribed by the same topological charge as the
Fermi-surface \cite{Volovik1991}.

The lines of zeroes generally have no topological stability
and an existence of the nodal lines can be prescribed only by the symmetry of
the superconducting state \cite{Grinevich1988}. The nontrivial classes  of the
superconductors, which symmetry  leads to the nodal lines in  symmetric
positions, are enlisted in Ref.\cite{VolovikGorkov1984}.
The symmetry violating external perturbation, such as impurities
or deviation of the crystal symmetry from the tetragonal one, destroys the line
zeroes \cite{Grinevich1988}. One could expect different types of behavior,
which should depend on the parameters of the system. Impurities can: (i)
produce the gap in the fermionic spectrum \cite{Pokrovsky1995}; (ii)
renormalize the exponent $\alpha$ in  $N(\omega)\propto \omega^{\alpha}$
\cite{Nersesyan1994}; (iii) lead to localization  \cite{Lee1993}; or (iv)
lead to the finite density of states \cite{Sun1995}.

 \section{ Fermions localized in the vortex core.}

\subsection{Fermion zero modes}

The energy $E(Q,p_z)$ of excitations localized in the axisymmetric vortex core
depends on the linear momentum $p_z$ along the symmetry axis and on the
discrete quantum number $Q$ \cite{Caroli1964}. $Q$ is the generalization of
the angular momentum (the eigenvalue of the generator of the "axial" symmetry
of the vortex \cite{SalomaaVolovik1987}). In the most symmetric vortices the
low-lying levels are
$$
E_0(p_z,Q)=Q\omega_0(p_z) ~~,
\eqno(3.1)
$$
where the interlevel distance  $\omega_0=\partial E /\partial Q \sim
\Delta^2/E_F$ is small compared to the gap amplitude $\Delta$ in the bulk
superconductor.

Depending   on the type of the vortex and on the pairing state the quantum
number $Q$ can be either integer or half of odd integer. In conventional
singly quantized vortices  (ie with winding number $n=1$) the quantum number
$Q$ is  half of odd integer, which means that excitations have small but
finite gap $(1/2)\omega_0(0)$. Since this gap is small, some perturbations
can modify the spectrum in such a way, that $E_0(p_z,\pm 1/2)$ can cross zero
energy as a function of  $p_z$. This is discussed for the vortices in
superconductors \cite{MakhlinVolovik1995a} and  $^3$He
\cite{Misirpashaev1995,Volovik1989}.

In doubly quantized ($n=2$) vortices and in some singly
quantized $^3$He-A vortices the quantum number $Q$ is integer
\cite{Misirpashaev1995}. This allows the fermionic states with
$Q=0$ which corresponds to the flat band: the energy $E_0(p_z,0)$ is zero for
all $p_z$. Such flat band in the vortex core has been first discussed in
\cite{KopninSalomaa1991}.

These details of the quasiparticle spectrum at low energy are important only
at very low temperature $T\sim \omega_0$. At higher temperatures $ T\gg
\omega_0$ the discreteness of the spectrum is sometimes not very important and
one can consider $Q$ as a continuous variable. In what follows we assume that
 $T_c\gg T\gg \omega_0$

The Eq.(3.1) shows that there is a  branch in the quasiparticle energy
spectrum which as a function of the "continuous" quantum number $Q$
crosses zero of energy. Such branches of the fermions localized on vortices or
other topological objects, which energy spectrum cross zero, are called
fermion zero modes. The existence of the fermion zero modes in a  background
the topological object is usually deduced by the application of certain index
theorems which relate the number of such modes to the topological charge of
the object, say, to the vortex winding number $n$. Originally such relation was
found for the  spectrum $E(p_z)$ of  fermions localized on strings in particle
physics (see Refs. \cite{JackiwRossi1980,WeinbergE1981,Hindmarsh1992}): the
number $N_{\rm zm}$ of the branches which as a function of $p_z$ cross the zero
level equals $n$.

For fermions in condensed matter vortices there is no such theorem, but a
similar theorem exists if one considers the spectrum as a function of the
quantum number $Q$. The number $N_{\rm zm}$ of the branches which as a function
of $Q$ cross the zero level is $N_{\rm zm}=2n$ \cite{Volovik1993a}. The
factor 2 corresponds to the spin degeneracy.

\subsection{Fermion zero modes in semiclassical approach}

The existence of fermionic zero modes does not
depend on the details of the system (provided that the interlevel distance
$\omega_0$ is small) and is completely defined by topology. So here we will
consider the simplest and best known case of an axisymmetric vortex in
superfluid or superconductor with $s$-wave pairing. The orbital quantum number
$Q$ is considered here as a continuous variable and so one can use the
quasiclassical approximation for the fermions localized in the vortex core.
The Bogoliubov Hamiltonian for the fermions with given spin projection is
a $2\times 2$ matrix
$$
{\cal H}=-{i\over m}\check \tau_3{\bf q}\cdot \vec\nabla  +
\check \tau_1 Re\Psi({\bf r})-\check
\tau_2 Im\Psi({\bf r})~.
\eqno(3.2)
$$
Here $\check \tau$ are the Bogoliubov-Nambu matrices in the particle-hole
space; ${\bf q}={\bf p}-{\hat {\bf z}}({\bf p}\cdot{\hat {\bf z}})$ is the
quasiparticle momentum in the transverse plane and
$$\Psi({\bf r})=e^{in\phi}\Delta(r) \eqno(3.3)
$$ is the gap function (order
parameter) in the axisymmetric vortex with winding number $n$.

The quantum numbers, which characterize the fermonic levels in this
approximation, are (i) the magnitude of transverse momentum  of quasiparticle
$q$, which  is related to the longitudinal projection of momentum
$q^2=p_F^2-p_z^2$,
(ii) the radial quantum number $n_r$, and (iii) the continuous impact parameter
$y={\hat {\bf z}}\cdot({\bf r}\times{\bf q})/q$. It is  related to the
angular momentum
$\hbar Q$ by $Q=qy$. Introducing the coordinate $x={\bf r}\cdot{\bf q}/q$
along ${\bf q}$, such that $r^2= x^2+y^2$, and assuming that in the important
regions one has $\vert y \vert\ll \vert x\vert$,  one obtains the
dependence of the
gap function  on $x$ and $y$:
$$
\Psi({\bf r})\approx \Delta(\vert x\vert ) ( sign(x) -in{y\over
{\vert x \vert}})~~,
\eqno(3.4)
$$
and the Hamiltonian:
$$
{\cal H}={\cal H}^{(0)}+{\cal H}^{(1)}~~,
$$
$$
{\cal H}^{(0)}(x)=-i \check \tau_3 {q\over
m}\nabla_x +\check \tau_1 \Delta(\vert x\vert ) sign(x) ~~,
~~{\cal H}^{(1)}(x,y)=n \check \tau_2 y{{\Delta(\vert x\vert ) }
    \over {\vert x \vert}}~~.
\eqno(3.5)
$$
The Hamiltonian ${\cal H}^{(0)}(x)$ is ``supersymmetric'' -
there is an operator $\check \tau_2$ which anticommutes with
${\cal H}^{(0)}(x)$, i.e. $\{ {\cal H}^{(0)},\tau_2\} =0$, and
it has an integrable eigenfunction corresponding to zero eigenvalue:
$$
{\cal H}^{(0)}\chi^{(0)}(x)=0~~,~~\chi^{(0)}(x) \propto
(\check \tau_0-\check \tau_2)\exp \left [
{-{m\over q}\int_0^{\vert x\vert }dr~\Delta(r)} \right ] ~~.
\eqno(3.6)
$$
Here $\check \tau_0$ is the diagonal $2\times 2$ matrix.

Using first order perturbation theory in
${\cal H}^{(1)}$ one obtains the lowest energy levels:
$$
E_0(Q,p_z)\equiv E_{n_r=0}(Q,p_z)\approx <0\vert {\cal H}^{(1)}\vert 0>=-ny
<{{\Delta(r)}\over r}>=-Qn\omega_0(p_z)~~,
\eqno(3.7)
$$
$$
\omega_0(p_z)={{1}\over q(p_z)}
       {{ \int_0^\infty dr\vert\chi^{(0)}(r)\vert ^2
{\Delta(r)}/r }\over
{\int_0^\infty dr\vert\chi^{(0)}(r)\vert^2 }} ~~.
\eqno(3.8)
$$

For $n=1$ this is the anomalous branch of the low-energy localized fermions
obtained in Ref. \cite{Caroli1964}. If the energy spectrum is considered as a
continuous function of $Q$, this anomalous branch crosses zero at $Q=0$.

\section{Vortices with broken symmetry.}

\subsection{Broken symmetry in $^3$He-B vortices. Point gap nodes in the
broken symmetry core.}

Four types of linear defects have been experimentally verified to exist in
$^3$He-B in a rotating container \cite{Krusius1993}. Among them there are two
types of $n=1$ vortices: in both of them the parity is broken within the core
and in one of them in addition the axial symmetry is broken
\cite{SalomaaVolovik1987}.

The discrete symmetry breaking leads to a new phenomenon (see Review
\cite{SalomaaVolovik1987}). In conventional vortices without broken parity
(Eq.(3.3)) the order parameter and thus the superfluid gap vanishes on the
vortex axis, ie  the vortex axis consists of the "normal" Fermi liquid with
the Fermi surface. In the broken symmetry core such Fermi surface on the axis
splits into the gap nodes which occupy some finite region of the vortex core,
as a result the singularity on the axis is smoothened: everywhere in the core
the system is in the superfluid state.  Now at each point
${\bf r}$ of such a smooth core there is one or several pairs of the point
nodes, ie the energy $E({\bf p},{\bf r})$ becomes zero at
${\bf p}=\pm p_a({\bf r}){\hat {\bf l}}_a({\bf r})$. Here the unit vector
$\hat  {\bf l}_a({\bf r})$ shows the direction of the $a$-th pair of the gap
nodes. There is an important relation between the topology of the $\hat {\bf
l}_a({\bf r})$-field and the winding number $n$ of the vortex:
$$n=  \sum _a \int _{over~core} {{dx\, dy} \over {4\pi }} \Biggl
({\hat {\bf l}}_a \cdot {{\partial {\hat {\bf l}}_a} \over {\partial x}} \times
{{\partial {\hat {\bf l}}_a} \over {\partial y}}\Biggr )
\hskip2mm . \eqno(4.1)$$
This means that,  for dissolving the singularity  on  the
axis of $n$-quantum vortex, the point gap nodes within the core should cover
the unit sphere $n$ times. In Sec.5 we shall use this equation for the
calculation of the spectral flow force on the moving vortex.

\subsection{Vortices and vortex sheet in $^3$He-A.}

The $^3$He-A is the superfluid with point gap nodes directed along the axis
${\hat {\bf l}}$ of the spontaneous orbital momentum of the Cooper pair.
According to
Eq.(4.1) this means that the vortices in $^3$He-A can be continuous with an
arbitrary size of the vortex core. Depending on the type of the vortex the core
region is limited by the size of the container, by spin-orbital coupling or by
external magnetic field.   The Eq.(4.1) is the generalization of the Mermin-Ho
relation first found  for the ${\hat {\bf l}}$ vector in
$^3$He-A \cite{MerminHo1976}.  Until now four different types of vorticity
have been observed and investigated in $^3$He-A in a rotating container
\cite{Parts1995}. Among them there are two types of continuous $n=2$
vortex lines, the singular $n=1$ vortex, and the
vortex sheet \cite{Parts1994}.   All of them have broken axial and discrete
symmetries.

The vortex sheet is reperesented by the soliton with accumulated vorticity.
Each vortex in this structure is continuous $n=1$ vortex, which can live
only   within the soliton; it represents some kind of the Bloch line which
separates the parts of the soliton having degenerate mirror-reflected
structures \cite{Heinila1995}. If the
soliton is present in the vessel, then under acceleration of the container the
"Bloch" vortices enter the soliton from the surface of container earlier than
the bulk vortices can form. As a result, instead of formation of the
conventional lattice of the isolated vortices, the vortex sheet grows and
finally occupies the whole container.  Such "Bloch" vortices within the domain
walls or grain boundaries in superconductors were discussed in
relation to the fractional magnetic flux \cite{Sigrist1989,Sigrist1995}.

\subsection{Vortices in heavy fermionic superconductors.}

Different possible types of the broken symmetry in the vortices were
discussed for the heavy fermionic superconductors (see
Review \cite{Lukyanchuk1995}). An interesting possibility is the dissociation
of the $n=1$ vortex into two $n=1/2$ vortices
\cite{Zhitomirsky1995}. More on $1/2$ vortices see in Conclusion.

\subsection{Broken symmetry within the cosmic strings.}

The spontaneous breaking of the  continuous symmetry in the cosmic strings has
been discussed by Witten \cite{Witten1985}. In this case the spontaneously
broken symmetry is the electromagnetic symmetry $U(1)$. This implies that the
core of the string is superconducting: there are nondissipative current states
in which the superconducting electric current along the vortex axis is
concentrated  within the core. In the $^3$He-B vortices with broken axial
symmetry such current states correspond to the twisting of the nonaxisymmetric
vortex core; this spiral structure of the vortex core has been observed in NMR
experiments \cite{Kondo1991}.

The strings in the electroweak model are similar to the vortices in $^3$He-A
(see \cite{VolovikVachaspati1996}). Electroweak strings with different
winding number $n$  have been considered in Ref. \cite{Achucarro1993}, they
correspond to the  $^3$He-A vortices with broken parity.

\subsection{Effect of the fermion zero modes on the core structure.}

The fermion zero modes on vortices can also trigger the instability
towards the symmetry breaking within the core.  The vortex-core
fermions which spectrum  $E(p_z)$ crosses zero as function of $p_z$, form
the $1D$ Fermi-liquid. The instabilities experienced by $1D$ fermionic system
at low $T$ lead  to the symmetry breaking in the vortex core.  In the case of
the electroweak strings this effect has been investigated in
\cite{Naculich1995,Liu1995}. The similar effect for the Abrikosov vortices in
superconductors has been discussed in \cite{MakhlinVolovik1995a}. Such
instability  is enhanced if instead of the $1D$ Fermi liquid there is a
flat band of fermions in the vortex core. Such a flat band can exist in
different types of symmetric vortices \cite{Misirpashaev1995}. It has been
also discussed for the  $0-\pi$ Josephson contacts in
high-temperature superconductors \cite{Hu1994}. The instability of the
flat band can lead to the breaking of the time inversion with formation of
the state with the big  ferromagnetic moment in the Josephson junction
\cite{Hu1994}.

The fermion zero modes, which spectrum  $E(Q)$ crosses zero as a  function of
$Q$, also influence the core structure at low  $T\ll T_c$ (though without
symmetry breaking).  An  anomalously large slope of the gap amplitude
$\Delta(r)$  at $r=0$  was analytically found in \cite{KramerPesch1974} and
numerically confirmed in \cite{GygiSchluter1991}.

\section{Spectral flow force on moving vortex}

The fermion zero modes also influence the dynamics of the vortex leading to
the special nondissipative force on the moving vortex \cite{Volovik1993a} which
is different from the conventional Magnus force. Let us first consider this
force for the simple case when the singularity on the vortex core is smoothed
by the formation of the region of the point gap nodes within the vortex core.

\subsection{Point gap nodes and axial anomaly.}

Near the $a$-th point gap node at ${\bf p}=\pm p_a{\hat {\bf l}}_a$ the
energy spectrum
of the Bogoliubov excitation has the following general form:
$$E^2({\bf p},{\bf r})=g^{ik}({\bf r})(p_i -qA_i({\bf r}))(p_k -qA_k({\bf
r}))~~.
\eqno(5.1)$$
Here ${\bf A}= p_a{\hat {\bf l}}_a$, and $q=\pm 1$. This energy corresponds
to that of the massless  particle, with ``electric''  charge $q$, which moves
in the ``electromagnetic'' field  ${\bf A}$ and in the gravity  field described
by the metric tensor $g^{ik}({\bf r})$ .

It also appears that these massless (gapless) quasiparticles in the vicinity of
the point gap nodes are chiral: like neutrino they are either left-handed  or
right-handed \cite{Volovik1992a}.  For such gapless chiral fermions the
phenomenon of the axial anomaly takes place \cite{Adler1969,BellJackiw1969}
which gives rise to  the production of, say, left particles from the vacuum in
the presence of the ``electric''  and ``magnetic''  fields. The
total number of left particles produced by these fields per unit time is given
by the Adler-Bell-Jackiw anomaly equation
$$
    {1\over {2\pi^2}}\int d^3r ~\partial_t {\bf A} \cdot
                  (\vec  \nabla \times {\bf A}  \, \, )  ~~.
\eqno(5.2)
$$
It is important that the left quasiparticle carries the linear momentum
$p_a{\hat {\bf l}}_a$ and   the equal momentum is carried by the left
quasihole. As a result the counterpart of the axial anomaly in condensed
matter leads to the net product of the quasiparticle linear momentum in the
time dependent texture \cite{Volovik1992a}:
$$
\partial_t  {\bf P}_{qp}=
     {1\over {2\pi^2}}\sum_a\int d^3r~ p_a\hat {\bf l}_a~(\partial_t {\bf A}
            \cdot (\vec  \nabla \times {\bf A}  \, \, ))  ~~.
\eqno(5.3)
$$
Since the total linear momentum is nevertheless conserved, this equation means
that momentum is transferred from the collective variables describing the
inhomogeneous superconductng state to the system of quasiparticles.
This is the linear-momentum anomaly related to the fermons in superfluids,
superconductors and ferromagnets, which was first identified
in $^3$He-A \cite{Volovik1986a,StoneGaitan1987}. It is caused by the spectral
flow of fermions through the point gap nodes.

\subsection{Momentum exchange between the  moving continuous vortex and the
heat bath. }

The spectral flow of fermions results in a curious exchange of the linear
momentum between the moving vortex and the heat bath. Let us consider first
the vortex with smoothened core. When the vortex moves with velocity ${\bf
v}_L$, the position of the gap nodes
$\hat {\bf l}_a$ texture becomes time dependent: $\hat {\bf l}_a=\hat {\bf
l}_a({\bf  r}-{\bf  v}_Lt)$. As a result the ``electric'' field arises
$$
{\bf E}=\partial_t {\bf A}=p_0\partial_t \hat {\bf l}_a=-p_0({\bf
v}_L\cdot\vec\nabla)
{\hat {\bf l}}_a  \, \, .
\eqno(5.4)
$$
Here we for simplicity consider the isotropic case with coordinate
independent magnitude of the momentum $p_a=p_0$.  According to the Eq.(5.3)
this leads to the production of quasiparticle momentum. This momentum is
absorbed by the heat bath which moves with the velocity ${\bf v}_n$ (for
superconductors this is the velocity of the crystal lattice, while for
superfluids the  ${\bf v}_n$ is the velocity of the normal component of the
liquid). The force  per unit length of the vortex acting on the vortex from
the heat bath is
$$
{\bf F}_{\rm spectral~flow} =
     {p_0^3~\over {2\pi^2}}\sum_a \int dx~dy~ {\hat {\bf l}}_a~
( (\vec  \nabla \times {\hat {\bf l}}_a)  \cdot (({\bf  v}_L-
{\bf  v}_n)\cdot\vec\nabla){\hat {\bf l}}_a)  \,   ~~,
\eqno(5.5)
$$

Simple transformation of this equation using the integration by parts gives
\cite{Volovik1992b}
$$
{\bf F}_{\rm spectral
~flow}={p_0^3\over 6\pi^2} ({\bf  v}_L-{\bf  v}_n)\times {\hat {\bf z}}
\sum_a \int
dx~dy~{\hat {\bf l}}_a \cdot (\partial_x {\hat {\bf l}}_a
\times \partial_y {\hat {\bf l}}_a)    ~~.
\eqno(5.6)
$$
Finally using the  Eq.(4.1) one obtains the anomaly contribution to the
force acting on the  vortex with winding number $n$:
$$
{\bf F}_{\rm spectral
~flow}=n \pi {p_0^3\over 3\pi^2} ~{\hat {\bf z}}\times ({\bf  v}_n-{\bf  v}_L)
~~,
\eqno(5.7)
$$

This force is reactive, i.e.
nondissipative, since it is even under time inversion (${\bf  v}\rightarrow
-{\bf  v}$, $n \rightarrow
-n$).   It does not depend on the details of the vortex
structure and is defined  by the winding number $n$ of the vortex and by
the magnitude of the momentum $p_0$ at which the gap node takes place. This
stresses the topological origin of this anomalous force.

Similar nondissipative force on the magnetic vortices and skyrmions in
magnets, first found in \cite{Nikiforov1983}, was also interprteted in terms
of the spectral flow of fermions in \cite{Volovik1986b,Stone1995}.

The Eq.(5.7) can be extended to case of singular core, ie to the limit of zero
radius of the region of gap nodes. In this limit the point gap nodes are
collected on the vortex axis to make there the gapless Fermi surface and thus
the normal Fermi liquid. The parameter $p_0$ in Eq.(5.7)
becomes the Fermi momentum $p_F$ of this Fermi surface, and the factor
$p_F^3/3\pi^2$ transforms to the particle density on the vortex axis
in accordance with suggestion made in \cite{Feigelman1995}. However, as was
shown by Kopnin \cite{Kopnin1996}, this result is model dependent.

\subsection{Spectral flow in  singular vortices}

The above analysis is valid only in the model when the core size
exceeeds the coherence length in superconductors, which is always true for
the $^3$He-A vortices. Here we calculate this spectral-flow force without
using of the smooth model for vortex core. In the quantum-mechanical
approach the effect occurs due  to the spectral flow of the $E(Q)$ zero modes
and is suppressed if the spectral flow is not allowed due to the discrete
nature of the quantum number
$Q$. If the spectral flow is not suppressed, the result for this force remains
the same (if one neglects the small possible deviation of $p_0$ from $p_F$)
and does not depend on the core structure.

We shall consider here the
conventional Abrikosov vortex in $s$-wave superconductor, but the result
remains the same for arbitrary vortices. It can be applied to vortex in the
$p$-wave $^3$He-B \cite{Kopnin1995} as well as to the vortex in the $d$-wave
superconductors: the complicated structure of the latter vortex
\cite{Volovik1993b,Berlinsky1995,Franz1995,Soininen1994,Ichioka1995} should
not influence the topological result. Note that for the $s$-wave superconductor
the result was first obtained long ago in Ref. \cite{KopninKravtsov1976} in a
rigorous microscopic theory.

If the vortex moves with the velocity ${\bf  v}_L$ with respect to the heat
bath, the coordinate ${\bf  r}$ is replaced by the ${\bf  r}-{\bf  v}_Lt$ and
the impact parameter  $y$ which enters the quasiparticle energy in Eq.(3.4)
shifts  with time. So the energy of zero mode becomes
$$
E(Q,p_z,t)=-n(y- {{\epsilon ({\bf  q})}\over q} t)q\omega_0(p_z)=-n(Q
-\epsilon ({\bf  q})t)~\omega_0(p_z)~~.\eqno(5.8)
$$
Here $\epsilon ({\bf  q})={\hat {\bf z}}\cdot(({\bf
v}_L-{\bf  v}_n)\times {\bf  q})$ acts on
fermions localized in the core in the same way that an
electric field acts on the fermions localized on a string in
relativistic quantum theory. The only difference is that under this
``electric'' field the  spectral flow occurs along $Q$ rather
than along $p_z$. Along this path the fermionic levels cross the zero
energy level at the rate
$$
\partial_t Q=\epsilon ({\bf  q})={\hat {\bf z}}\cdot(({\bf
v}_L-{\bf  v}_n)\times {\bf  q})~~.
\eqno(5.9)
$$
This leads to the quasiparticle momentum transfer from the vacuum (from
the levels below zero) along the anomalous branch into the heat bath.
This occurs at the rate
$$\partial_t{\bf  P}=\sum_{{\bf  p}} {\bf  p}~\partial_t Q= {1\over 2}N_{\rm
zm}\int_{-p_F}^{p_F} {{dp_z}\over{2\pi}}
\int_0^{2\pi}{{d\phi}\over{2\pi}}~{\bf  q}
\epsilon ({\bf  q})=\pi n {{p_F^3}\over {3\pi^2}}{\hat {\bf z}}\times({\bf
v}_n-{\bf  v}_L),~~
\eqno(5.10)
$$
where the factor ${1\over 2}$ compensates the double counting of particles
and holes,
and we used the index theorem that the number of anomalous
branches (fermion zero modes) is related to the winding number: $N_{\rm
zm}=2n$. The Eq.(5.10) is the result, which was obtained in Sec.5.2 within
the model of smooth core.

Here it is implied that all the quasiparticles, created from the negative
levels of the vacuum state, finally become the part of the normal
component ouside the core. This means that there is nearly reversible transfer
of linear momentum from the core fermions to the heat bath, which is valid only
in the limit of large scattering rate: $\omega_0\tau\ll 1$, where $\tau$ is the
lifetime of the fermion on the level $Q$. The irreversibility of the momentum
transfer leads to the small friction force ${\bf F}_{\rm friction}\propto
\omega_0\tau ({\bf  v}_n-{\bf  v}_L)$.  The condition  $\omega_0\tau\ll 1$,
which states that the interlevel distance on the zero mode branch is small
compared to the life time of the level, is the crucial requirement for
spectral flow to exist.  In the opposite limit $\omega_0\tau\gg 1$ the spectral
flow is suppressed and the corresponding spectral flow force is exponentially
small \cite{KopninVolovik1995}.

\subsection{Magnus and spectral-flow forces}

The reactive force ${\bf F}_{spectral~flow}$ from the heat bath on the
moving vortex is the consequence of the reversible flux of momentum from
the vortex into the region near the axis, i.e. into the core region.
Within the core the linear momentum of the vortex transforms to the
linear momentum of the fermions in the heat bath when the fermionic
levels on anomalous branches cross zero.

The Magnus force comes from the flux of linear
momentum from the vortex to infinity. The topological origin of this force
is similar to the
Aharonov-Bohm effect. This force can be obtained as a sum of
the forces acting on the individual particles according to the equation
$$
\partial_t{\bf p}= (\vec\nabla\times{\bf v}_s)\times{\bf p} ,~~
\eqno(5.11)
$$
where ${\bf p}$ is the particle
momentum and the vorticity
$\vec\nabla\times{\bf v}_s=n{\pi\hbar\over m}\delta_2({\bf r})$
is concentrated in a thin tube (vortex core). The  force on the vortex is
$$
- \sum_{{\bf p}} n({\bf p}) \partial_t{\bf p}=-(\int
d^2r (\vec\nabla\times{\bf v}_s)) \times ~\int {{d^3p} \over
{4\pi^3}}~ {\bf p}n({\bf p})= n\pi {\hat {\bf z}}\times  {\bf j} ,~~
\eqno(5.12)
$$
Here $n({\bf p})$ is the particle distribution function and ${\bf j}$ is the
total particle current in the frame of the moving vortex. This force can be
expressed through the superfluid ${\bf v}_s$  and normal ${\bf v}_n$
velocities as a sum of the Magnus and Iordanskii forces
\cite{Iordanskii1964,Iordanskii1965}
$$
 {\bf F}_{\rm Magnus} + {\bf F}_{\rm Iordanskii} = n\pi {\rho_s\over
m}{\hat {\bf z}}\times  ({\bf v}_L-{\bf v}_s) +  n\pi {\rho_n\over
m}{\hat {\bf z}}\times  ({\bf v}_L-{\bf v}_n).~~
\eqno(5.13)
$$
On the connection between the Iordanskii force and the Aharonov-Bohm effect has
been pointed out in Ref. \cite{Sonin1975}. Similar effect for the
spinning cosmic string  was discussed in \cite{Harari1988}.

The  two forces in Eq.(5.13) exist both in Bose and Fermi superfluids, as
distinct from the spectral-flow force which originates from the fermion zero
modes in the core. The latter is temperature independent
\cite{MakhlinMisirpashaev1995}, if the condition $\omega_0\tau\ll 1$ is
satisfied. When $\omega_0\tau $ is not small, then far from $T_c$ the
spectral-flow force can be approximated by the following  equation
\cite{KopninVolovik1995,KopninLopatin1995}:
$$
{\bf F}_{\rm spectral
~flow}=n \pi {p_0^3\over 3\pi^2} {1\over 1+\omega_0^2\tau^2}  {\hat {\bf
z}}\times ({\bf  v}_n-{\bf  v}_L) ~~.
\eqno(5.14)
$$

The balance of three nondissipative forces in Eqs.(5.14-15) acting on the
vortex and of the friction force, which is $\propto {\bf  v}_n-{\bf  v}_L$,
governs the low-frequency dynamics of the vortex.
The spectral-flow force disappears in the limit
$\omega_0\tau\gg 1$. On the other hand, if the  spectral-flow force has its
largest value, it nearly compensates the Magnus force, since  the total mass
density $\rho=\rho_s+\rho_n$, which enters Magnus+Iordanskii
force,  is very close to the parameter  $m p_F^3/ (3\pi^2)$, which enters the
spectral-flow force. That is why the spectral flow  within the core plays a
very important part in the vortex dynamics and can lead to the change of the
sign of the Hall effect \cite{vanOtterlo1995,Feigelman1995}.

Note that the spectral-flow parameter $p_0^3/ (3\pi^2)$ is close to $ \rho/m$
only in the weak coupling limit, where the difference between these parameters
is coursed by a tiny asymmetry between particles and holes near the Fermi
surface. If the interaction which leads to the Cooper pairing is strong
enough,  $p_0^3/ (3\pi^2)$ can essentially deviate from
$\rho/m$ and even become zero \cite{Volovik1992a}. In the latter case the
spectral-flow force disappears completely. The situation is exactly the same as
for the problem of the intrinsic angular momentum in $^3$He-A, which
magnitude $L_0$ is essentially modified by the spectral flow: instead of the
nominal value $L_0=(\hbar/2) \rho/m$ in the Bose liquid with the symmetry
of $^3$He-A one has $L_0=(\hbar/2)[\rho/m - p_0^3/ (3\pi^2)]$ in the
Fermi liquid.

In $^3$He-B the spectral-flow  contribution to
the nondissipative force has been measured
\cite{Bevan1995}: the experimental temperature dependence follows the
theoretical one found in \cite{KopninVolovik1995,Kopnin1995}.

The spectral-flow force has been also calculated for Josephson vortices in
SNS junction \cite{MakhlinVolovik1995b}: it almost cancels the
Magnus force, again due to approximate particle-hole symmetry. This possibly
provides an explanation for the experimental evidence of the negligible Magnus
force in 2D Josephson junction arrays
\cite{vanZant1992,Lachenmann1994,vanZant1995}.

\section{Conclusions}

The fermion zero modes in the bulk condensed matter and within the topological
objects are of great importance at low $T$, where they lead to different
anomalies. In particle phisics the anomaly related to the spectral flow could
give rise to the baryogenesis at the early stage of the expanding universe
\cite{Turok1995}. In condensed matter the similar phenomenon is responsible
for the ``momentogenesis´´ - production of the linear momentum during the
vortex motion, which results in additional force acting on the vortex line.

An open question is the behavior of the fermions in the presence of new
topological object in high-temperature superconductor -- the  half-quantum
vortex, which has been discussed in \cite{Geshkenbein1987} and  recently
observed in \cite{Kirtley1996}. One may expect different interesting phenomena
related to such vortices. For the case of superfluid $^3$He-A, where such
vortices can also exist \cite{VolovikMineev1976b} but still have not been
identified, they are the counterpart of the Alice string in some models of
particle physics. The person travelling around Alice string finds himself in
the mirror reflected world \cite{Schwarz1982}.  In superfluid
$^3$He-A, the analogous effect is the reversal of the spin of the
quasiparticle upon circling the 1/2 vortex. The
peculiar Aharonov-Bohm effect for 1/2 vortices has been discussed for $^3$He-A
\cite{Khazan1985,SalomaaVolovik1987} and
modified for the Alice string \cite{March-Russel1992,Davis1994}.

The 1/2 vortex observed in high-temperature superconductor
carries 1/2 of the magnetic flux quantum $\Phi_0=h/2e$  \cite{Kirtley1996}. In
principle the objects with the fractional flux below
$\Phi_0/2$ are also possible \cite{VolovikGorkov1984,Sigrist1989,Sigrist1995}.
They can arise if the time inversion symmetry is broken. The behavior of the
fermions in the background of these and other extended topological
objects in superconductors, superfluids, magnets and Quantum Hall systems is
of primary interest.

\vskip 0.5 truecm
\noindent {\it Acknowledgements:}

This work was supported through the ROTA co-operation plan of the Finnish
Academy and the Russian Academy of Sciences and by the Russian Foundation
for Fundamental Sciences, Grant 96-02-16072.

\vfill
\eject


\vfill
\eject

\end{document}